\documentclass[aps,prl,superscriptaddress,twocolumn,showpacs]{revtex4}

\usepackage{psfig}

\begin{document}

\title{Complete Two Loop Bosonic Contributions to the Muon Lifetime \\
 in the Standard Model}

\author{M. Awramik}

\affiliation{Institut f\"ur Theoretische Teilchenhysik, 
Universit\"at Karlsruhe,
D-76128 Karlsruhe, Germany}

\affiliation{Department of Field Theory and Particle Physics,
Insitute of Physics, University of Silesia,
Uniwersytecka 4, PL-40007 Katowice, Poland}

\author{M. Czakon}

\affiliation{Institut f\"ur Theoretische Physik, 
Universit\"at Karlsruhe,
D-76128 Karlsruhe, Germany}

\affiliation{Department of Field Theory and Particle Physics,
Insitute of Physics, University of Silesia,
Uniwersytecka 4, PL-40007 Katowice, Poland}

\begin{abstract}
The last missing correction to the muon lifetime in the Standard Model
at ${\cal O}(\alpha^2)$ coming from gauge and  Higgs boson loops is
presented. The associated contribution to the parameter $\Delta r$ in
the on-shell scheme ranges from $6 \times 10^{-5}$ to $-4 \times
10^{-5}$ for Higgs boson masses from 100~GeV to 1~TeV. This result
translates into a shift of the $W$ boson mass which does not exceed
$\pm 1$~MeV in the same range and amounts in particular to
approximately $- 0.8$~MeV for a 115~GeV Higgs boson.
\end{abstract}


\maketitle

The muon decay lifetime ($\tau_\mu$) has long been used as an input
parameter for high precision predictions of the Standard Model
(SM). It allows for an indirect determination of the mass of the $W$
boson ($M_W$), which suffers currently from a large experimental error
of 39~MeV \cite{Hagiwara:pw}, one order of magnitude worse than that
of the $Z$ boson mass ($M_Z$). A reduction of this error by the Large
Hadron Collider (LHC) to 15~MeV \cite{lhctdr} and by a future linear
collider to 6~MeV \cite{tesla-tdr} would provide a stringent test of
the SM by confronting the theoretical prediction with the experimental
value.

The extraction of $M_W$ with an accuracy matching that of next
experiments, {\it i.e.} at the level of a few MeV necessitates
radiative corrections beyond one loop order. Large two loop
contributions from fermionic loops have been calculated in
\cite{Freitas:2000gg}.  The current prediction is affected by two
types of uncertainties. First, apart from the still unknown Higgs
boson mass, two input parameters introduce large errors. The current
knowledge of the top quark mass results in an error of about 30~MeV
\cite{Freitas:2002ja}, which should be reduced by LHC to 10~MeV and by
a linear collider even down to 1.2~MeV. The inaccuracy of the
knowledge of the running of the fine structure constant up to the
$M_Z$ scale, $\Delta \alpha(M_Z)$, introduces a further $6.5$~MeV
error. Second, several higher order corrections are unknown, of which
only one at the ${\cal O}(\alpha^2)$ order. This lacking contribution
comes from diagrams with no closed fermion loops. It has been
previously estimated to be of the order of the square of the one loop
bosonic correction \cite{Freitas:2000gg}. This,  however, is
unjustified since both have a different dependence on the Higgs boson
mass (logarithmic vs. quadratic). In this Letter, the question
of the exact size of this contribution is finally settled.

The muon decay is naturally described in the language of effective
field theory. The process' dynamics are driven by the leptons, the
five light quarks, and the two massless gauge bosons. Even then the
momentum scale set by the muon mass leads to a strong decoupling of
the $\tau$ lepton and of the heavier of the five quarks
\cite{vanRitbergen:1998hn}.  The heavy particles $W$, $Z$, the Higgs
boson and the top quark generate point interactions. A general
effective lagrangian assumes the form
\begin{equation}
  \label{Leff}
  {\cal L}_{\text{eff}} = {\cal L}_{\text{QED}}
  +{\cal L}^{(5)}_{\text{QCD}}+\sum_{n,i} 
  \frac{{\cal C}_n^i}{(M_W^2)^n} {\cal O}_n^i,
\end{equation}
where ${\cal L}_{QED}$ and ${\cal L}^{(5)}_{QCD}$ are the bare QED
and the five flavour QCD lagrangians, ${\cal O}^i_n$ are composite
operators of the light fields and ${\cal C}^i_n$ are the respective
dimensionless matching coefficients, obtained by comparing the Green
functions of the full theory with those of the effective one. The mass
of the $W$ boson has been chosen as the heavy scale.

Due to the left-handed nature of the charged current in the SM, the
only operator of dimension six relevant to muon decay is the
four-fermion interaction of the Fermi model
\begin{equation}
  \label{OF}
  {\cal O}_{F} = \bar{e} \gamma^\alpha (1-\gamma_5) \mu
  \otimes \bar{\nu_\mu} \gamma_\alpha (1-\gamma_5) \nu_e,
\end{equation}
with $\mu$, $e$, $\nu_\mu$, $\nu_e$ denoting the spinors of the muon,
electron and their neutrinos. Here the so called charge-conserving
form has been used (Fierz transformed SM amplitude). The respective
matching coefficient is traditionally parametrised as
\begin{equation}
  \label{CF}
  {\cal C}_{F} = M_W^2 \frac{G_F}{\sqrt{2}} = \frac{\pi
  \alpha}{2 s_W^2}(1+\Delta r),
\end{equation}
where $G_F$ is known as the Fermi constant, whereas $\Delta r$
represents the higher order corrections and vanishes at the Born
level. The one loop corrections in the effective theory given by
Eqs.~\ref{Leff},~\ref{OF} and \ref{CF} have been calculated in
\cite{Berman:1958ti} and \cite{Kinoshita:1958ru}, while the two loop
in \cite{vanRitbergen:1998yd}.

\begin{figure}
\psfig{file=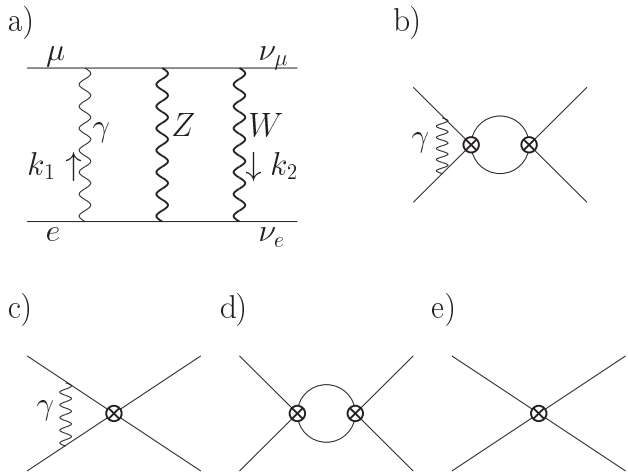,width=8.6cm}
\caption{\label{expansions}
A typical muon decay diagram a) and the contributions to its large
mass expansion according to the momenta b) $k_1$-soft, $k_2$-soft; c)
soft-hard; d) hard-soft; e) hard-hard.}
\end{figure}

The framework for construction of effective lagrangians together with
a proof has been given in \cite{Gorishnii:dd}. It is based on
expansions of individual Feynman diagrams. The method is presented on
the example of a specific graph contributing to bosonic corrections to
muon decay at two loop order in Fig.~\ref{expansions}. The terms in
the large mass expansion can be classified according to the scale of
the loop momenta. When both momenta are ``soft'' ($\ll M_W$), as in
b), the propagators of the $W$ and $Z$ bosons are expanded leading to
a correction of order $\alpha/M_W^4$ in the effective theory. For one
momentum ``soft'' and one ``hard'' ($\sim M_W$), as in c) and d),
corrections of either order, $\alpha/M_W^2$ or $1/M_W^4$ in the
effective theory, are generated. The contribution to the matching
coefficient comes only from the region where both momenta are hard, as
in e). In this case all of the light particle masses and momenta
should be put to zero. By these arguments it can be shown that $\Delta
r$ can be obtained by simply taking the sum of all the diagrams and
putting all external momenta and light masses to zero. The procedure
should generate no spurious infrared divergences, while the physical
divergences connected with the photon should be contained in the
corrections of the effective theory. As known the Fermi theory
corrections are finite, therefore the $\Delta r$ correction obtained
as above should also be finite.

Previous calculations of $\Delta r$ have been based on a different
method of factorisation originally devised in \cite{Sirlin:1980nh}.
This procedure consists of subtracting from the infrared divergent SM
diagrams the respective Fermi theory diagrams in Pauli-Villars
regularisation.  The difference is well defined in the limit of zero
light masses and external momenta. It turns out, however, that the
QED Ward identity, which is responsible for the finiteness of the
corrections in the Fermi theory, implies in this case the vanishing of
the sum of the subtracted diagrams. This proves that both procedures
are equivalent.


The evaluation of two loop corrections to a four-fermion process
requires the full second order renormalisation of the SM lagrangian in
all but the Higgs sector, where first order suffices. The comparison
with experiment imposes the use of on-shell parameters for the final
result. Throughout this work the on-shell scheme was used, with a
procedure similar to the one described in \cite{Freitas:2002ja}. The
only substantial difference concerns the treatment of tadpoles.

It is known that gauge invariance of mass counter-terms requires
inclusion of tadpoles \cite{appelquist,tadpole} (at the two loop level
this has been explicitely shown in \cite{Jegerlehner:2001fb}). In this
case, however, one cannot use  one-particle-irreducible (1PI) Green
functions. In order to have gauge invariant counter-terms and 1PI
Green functions only, a special procedure was designed. An additional
renormalisation constant for the bare vacuum expectation value $v_0$,
denoted $Z_v$, has been introduced and explicitely split from the bare
masses
\begin{eqnarray}
  v_0 & \longrightarrow & v_0 Z_v^{1/2}, \\ 
  M^0_{W,Z} & \longrightarrow & M^0_{W,Z} Z_v^{1/2}.
\end{eqnarray}
The term linear in the Higgs field $H$ in the lagrangian
\begin{equation}
  T^0 H^0 = \frac{M^0_W s^0_W}{e_0} (M^0_H)^2 Z_v^{1/2} (Z_v-1) H^0,
\end{equation}
is then used to determine $Z_v$, through the requirement that tadpoles
are canceled. It can be proved \cite{appelquist,awramik} that the bare
masses are gauge invariant in this case (an equivalent procedure which
makes use of the effective potential has been used in \cite{bochkarev}).


The calculation of the two loop bosonic contributions to muon decay
was performed by means of a completely automated system. The diagram
generation stage was done by the C++ library {\bf DiaGen}
\cite{czakon}. The tensor reduction of two loop propagator diagrams
was accomplished with the algorithm described in \cite{Weiglein:hd},
whereas vacuum diagrams were treated with integration by parts
identities \cite{Chetyrkin:qh}. For algebraic manipulations, the
program {\bf FORM} \cite{Vermaseren:2000nd} was used. The two loop
two-point integrals were numerically evaluated with single integral
representations of the package {\bf S2LSE}
\cite{Bauberger:1994by}. The latter was modified for quadruple
precision, which was needed due to large cancellations (independent
terms grow as $M_H^8$, while the result behaves as $M_H^2$).

The size of the software required several tests. The following
algebraic checks were performed
\begin{itemize}
\item ultraviolet and infrared finiteness, by cancellation of poles in
  dimensional regularisation,
\item gauge invariance, by independence of the three gauge parameters
  of the general $R_\xi$ gauge for the SM,
\item Slavnov-Taylor identities for two-point functions, as given in
  \cite{Weiglein:hd}, both for on-shell integrals and by expansion in
  the external momentum to second order.
\end{itemize}
Several numerical tests were also done
\begin{itemize}
  \item all of the master integrals were evaluated independently by
    means of deep mass difference and large mass expansions,
  \item each of the two-point on-shell diagrams was calculated
    separately with the help of small-momentum and different large-mass
    expansions,
  \item the result of \cite{Jegerlehner:2001fb} for the $W$ and $Z$
    mass counter-terms was reproduced to precision dictated by the
    order of the expansions contained therein.
\end{itemize}
A detailed description of the methods used is relegated to a subsequent
publication \cite{awramik}.

The variation of the bosonic contribution with $M_H$ subtracted at
$M_H = 100$~GeV has also been compared with the one presented in
\cite{Freitas:2002ja}. A discrepancy which grows up to $\sim 16\%$
for $M_H = 1$~TeV, has been found.

All the numerical values have been obtained for the following coupling
and mass parameters \cite{Hagiwara:pw}
\begin{eqnarray}
  \label{parameters}
  && \alpha^{-1} = 137.03599976(50), \\
  && G_F = 1.16639(1) \times 10^{-5} \mbox{ GeV}^{-2}, \nonumber \\
  && M_W = 80.423(39) \mbox{ GeV}, \nonumber \\
  && M_Z = 91.1876(21) \mbox{ GeV}. \nonumber
\end{eqnarray}


\begin{figure}
\psfig{figure=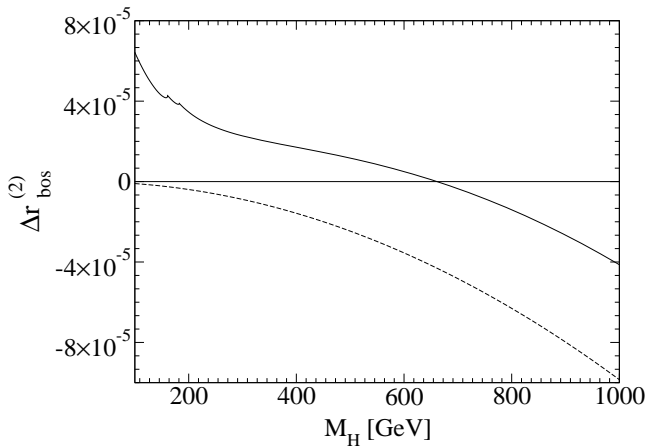,width=8.6cm,angle=270}
\caption{\label{dr2OS} 
$\Delta r$ as function of the Higgs boson mass (solid line) and the
leading term in the large Higgs mass expansion (dashed line).}
\end{figure}

The two loop bosonic correction to $\Delta r$, denoted $\Delta
r^{(2)}_{bos}$, is shown in Fig.~\ref{dr2OS} as function of the Higgs
boson mass $M_H$. Its behaviour is very smooth apart from the range
between 150~GeV and 190~GeV, where two peaks are located. Their
positions at $\sim 161$~GeV and $\sim 182$~GeV correspond to the $W$
and $Z$ pair production thresholds introduced through the on-shell
definition of the Higgs boson mass.

It is interesting to study the correction for large
values of $M_H$. In fact, two different leading $M_H^2$ terms have been given
in \cite{Halzen:1991ik} and \cite{Jegerlehner:1991ed}, both based
on the same results of \cite{vanderBij:1983bw}. In this last work a
renormalisation scheme has been used in which $M_H^2$ terms occur only
in two-point functions. The leading behaviour can then be obtained
from the formula
\begin{equation}
  \Delta r^{(2)}_{\text{Higgs}} = 2 \frac{\delta e}{e} + \frac{\Delta M_W^2}{M_W^2}
  +\frac{c_W^2}{s_W^2} \left( \frac{\Delta M_Z^2}{M_Z^2} 
  - \frac{\Delta M_W^2}{M_W^2} -\Delta \rho \right),
\end{equation}
where $\delta e$ is the charge renormalisation and $\Delta M_{W,Z}$
represent the shift of the $W$ and $Z$ self energies from zero
external momentum up to the on-shell value. $\Delta \rho$ has been
introduced in \cite{Ross:1975fq}. With the help of
\cite{vanderBij:1983bw}, the result reads
\begin{eqnarray}
  \label{leading}
  \Delta r^{(2)}_{\text{Higgs}} &=& \left( \frac{\alpha}{4 \pi s_W^2} \right)^2
  \frac{M_H^2}{8 M_W^2} \\ &\times&
  \left( 9 \sqrt{3} \mbox{ Cl}_2 (\frac{\pi}{3}) + \frac{49}{72} 
  - \frac{11 \pi \sqrt{3}}{4} - \frac{25 \pi^2}{108} \right), \nonumber
\end{eqnarray}
where Cl$_2$ is the Clausen function and Cl$_2(\pi /3) =
1.0149416$. Eq.~\ref{leading} is in agreement with
\cite{Halzen:1991ik}, whereas in \cite{Jegerlehner:1991ed} the charge
renormalisation has not been included. The leading Higgs mass
dependence is depicted in Fig.~\ref{dr2OS}. The difference between the
full result and Eq.~\ref{leading} for values below 1~TeV shows the
importance of subleading corrections. It has been checked that for
$M_H > 5$~TeV both functions agree very well.


\begin{figure}
\psfig{figure=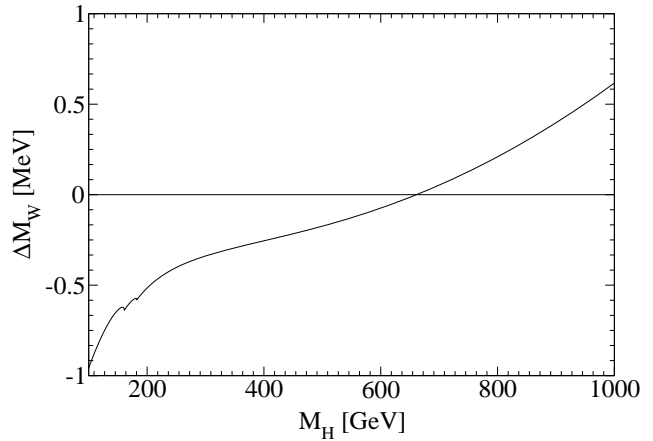,width=8.6cm,angle=270}
\caption{\label{WMassShift} 
$W$ boson mass shift generated by two loop bosonic corrections.}
\end{figure}

With the help of $\Delta r$ the $W$ boson mass can be estimated by
inverting Eq.~\ref{CF}
\begin{equation}
  \label{MW}
  M_W = M_Z \sqrt{\frac{1}{2}+\sqrt{\frac{1}{4} -\frac{\pi
\alpha}{\sqrt{2} G_F M_Z^2}(1+\Delta r)}}.
\end{equation}
This equation should be solved recursively, since $\Delta r$ also
depends on $M_W$. In the case of small corrections one can, however,
obtain the additional mass shift by means of an expansion. For
parameters' values given in Eq.~\ref{parameters}, a second order
Taylor series yields
\begin{equation}
  \label{DMWEQ}
  \Delta M_W = -(1.491+1.779 \; \overline{\Delta r})\times 10^{4}
  \Delta r^{(2)}_{bos} \mbox{ [MeV]},
\end{equation}
where $\overline{\Delta r}$ represents the dominating correction,
which is composed of one loop electroweak, two loop fermionic and
${\cal O}(\alpha \alpha_S)$ QCD contributions. Since $\overline{\Delta r}$ 
does not exceed 5\% \cite{Freitas:2002ja}, the third order in the
Taylor series would contribute less than a percent which is comparable to
the inaccuracy induced by the dependence of $\Delta r^{(2)}_{bos}$ on the
$W$ mass.

The $W$ mass shift, Eq.~\ref{DMWEQ}, with $\overline{\Delta r}$
neglected, is plotted in Fig.~\ref{WMassShift}. The dropped term
implies an error of less than 6\%, which is, however, completely
negligible in view of the size of the total effect, which does never
exceed $\pm 1$~MeV.


In conclusion, the complete ${\cal O}(\alpha^2)$ bosonic contributions
to muon decay have been calculated in the on-shell scheme, and shown
to lead to a small $W$ boson mass shift below 1~MeV. A framework for
evaluation of $\Delta r$ which does not explicitely refer to the QED
corrections in the Fermi theory has been presented and its equivalence
to the method based on subtraction of Pauli-Villars regulated diagrams
demonstrated. A procedure of obtaining gauge invariant mass
counter-terms without explicitely including tadpole graphs has been
developed.


The authors would like to thank K. Chetyrkin, O. Veretin and
A. Onishchenko for fruitful discussions and A. Freitas for providing
the results of \cite{Freitas:2002ja}.  M. A. would like to thank the
``Marie Curie Programme'' of the European Commission for a stipend.
M. C. would like to thank the Alexander von Humboldt foundation for
fellowship. This work was supported in part by the European
Community's Human Potential Programme under contract
HPRN-CT-2000-00149 Physics at Colliders, and by the KBN Grants
5P03B09320 and 2P03B05418.


\end{document}